# Structural Feature in Dynamical Processes Accelerated Transition State Calculations

Hongsheng Cai[1,†], Guoyuan Liu[1,†], Peiqi Qiu[1], and Guangfu Luo[1,2,*]

[1]Department of Materials Science and Engineering, Southern University of Science and Technology, Shenzhen, Guangdong 518055, China.

[2]Guangdong Provincial Key Laboratory of Computational Science and Material Design, Southern University of Science and Technology, Shenzhen, Guangdong 518055, China.

[†]These authors contribue equally to this work.
[*]E-mail: luogf@sustech.edu.cn

**Abstract**

Minimum energy path (MEP) search is a vital but often very time-consuming method to predict the transition states of versatile dynamic processes in chemistry, physics, and materials science. In this study, we reveal that the chemical bond lengths in the MEP structures, including those directly involved in the dynamical processes, largely resemble those in the stable initial and final states. Based on this discovery, we propose an adaptive semi-rigid body approximation (ASRBA) to construct a physically reasonable initial guess for the MEP structures, which can be further optimized by the nudged elastic band method. Examination of several distinct dynamical processes in bulk, on crystal surface, and through two-dimensional system show that the transition state calculations based on the ASRBA results are robust and significantly faster than those based on the popular linear interpolation and image-dependent pair potential methods.

## I. INTRODUCTION

Transition state calculation is one major method to quantitatively reveal the atomistic mechanisms of dynamical processes in chemistry, physics, and materials science.[1] It lies in searching the minimum energy path (MEP) in high dimensional space and is an intrinsically challenging task. For transition state calculations in solids, the climbing-image nudged elastic band (cNEB)[2] approach is widely used, which requires structures of the initial and final states of a dynamical process, together with an initial guess of the MEP structures between them. Therefore, the accuracy of the initial guess of MEP structures largely determines the computational cost of a transition state calculation.

Up to now, linear interpolation (LI), linear synchronous transit (LST),[3] and image-dependent pair potential (IDPP)[4] are three major methods of generating an initial guess of the MEP structures. For the LI method, the coordinates of atom $i$ in the $n$th initial MEP structure are determined by Eq. (1)

$$\widetilde{r_i^n} = \left(1 - \frac{n}{N}\right)r_i^\alpha + \frac{n}{N}r_i^\beta, \quad n = 1, \ldots, N - 1, \tag{1}$$

where $r_i^\alpha$ and $r_i^\beta$ are atomic coordinates in the initial state $\alpha$ and final state $\beta$, respectively; $N - 1$ denotes the total number of MEP structures between the initial and final states. Therefore, the LI method takes only atomic coordinates into consideration and assumes a translational trajectory, which can easily lead to atom overlapping (too close atomic pair distance) for curved ones.[5]

Instead of linear interpolation of atomic coordinates, the LST and IDPP methods linearly interpolate the atomic pair distances between the initial and final states, as shown in Eq. (2)

$$\widetilde{d_{ij}^n} = \left(1 - \frac{n}{N}\right)d_{ij}^\alpha + \frac{n}{N}d_{ij}^\beta, \quad n = 1, \ldots, N - 1, \tag{2}$$

where $d_{ij}^\alpha$ and $d_{ij}^\beta$ represent the pair distances between atom $i$ and $j$ for the initial and final states, respectively. Because the number of pair distances is usually more than the total atomic degrees of freedom, $N_{atom}*(N_{atom}-1)/2$ versus $3N_{atom}$ for a total number of $N_{atom}$ atoms, Eq. (2) cannot be fully complied in most situations. Therefore, the objective functions in Eqs. (3) and (4) are constructed for the LST and IDPP methods, respectively, for minimization.

$$S_{LST}^n \equiv \sum_{i>j}[\widetilde{d_{ij}^n} - |r_i^n - r_j^n|]^2/\widetilde{d_{ij}^n}^4 + 10^{-6}\sum_i |r_i^n - \widetilde{r_i^n}|^2 \tag{3}$$

$$S_{IDPP}^n \equiv \sum_{i>j}[\widetilde{d_{ij}^n} - |r_i^n - r_j^n|]^2/\widetilde{d_{ij}^n}^4 \tag{4}$$

The LST and IDPP methods efficiently avoid atom overlapping and naturally describe rotational dynamics because of the rotational invariance nature of Eq. (2). One critical difference between the LST and IDPP method is that the IDPP method further utilizes the NEB method[6] to maintain an equal distance between the intermediate structures based on the gradient of Eq. (4), i.e. $-\nabla S^n_{IDPP}$, and thus avoids a discontinuous trajectory that LST could encounter.

To sum up, the LI, LST, and IDPP methods are based directly on the atomic coordinates and no physical or chemical properties are explicitly taken into consideration. To demonstrate the deficiencies of these methods, we propose a simplified dynamical process with one distant red atom passing through a blue atomic chain, as shown in Fig. 1a. In this process, the LI method predicts intermediate structures without considering the position of atomic chain and leads to very close atomic distances (Fig. 1b), which can result in the failure of subsequent transition state calculation. Because the initial and final states are symmetrical relative to the atomic chain, the atomic pair distances are equal between the initial and final states. This results in identical pair distances for all the intermediate structures according to Eq. (2) and thus the intermediate structures are forced to crowd around the initial and final states for the LST method (Fig. 1c). The IDPP method avoids the discontinuous trajectory issue related to the LST method by the NEB calculation, but it gives the same result for diffusion atoms with different sizes (Fig. 1d). In this study, we investigate the structural feature along the MEP and propose an adaptive semi-rigid body approximation (ASRBA) method to construct physically reasonable initial guess of the MEP structures. Application of the ASRBA method in several distinct dynamical processes clearly demonstrates its robustness and noticeable acceleration of the transition state calculations.

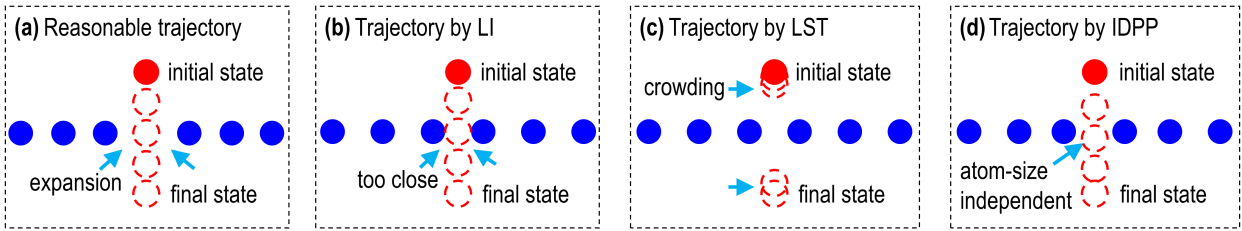

**FIG. 1.** Schematic processes demonstrating the deficiencies of the LI, LST, and IDPP methods. (a) Reasonable trajectory exhibits expansion in the atomic chain; (b) the LI method generates linear pathway with atom overlapping for the middle structure; (c) the LST method generates discontinuous trajectory; (d) the IDPP method generates atom-size independent intermediate structures.

## II. METHODS

The intermediated structures of the LI and IDPP method were obtained using the convasp[7] and the diffusion module in pymatgen,[8,9] respectively. The ASRBA method was implemented in the pymatgen. The cNEB calculation[2,10,11] based on DFT were conducted using the Vienna Ab initio Simulation Package,[12] together with the PBE exchange-correlation functional.[13] A plane-wave energy cutoff of 299, 299, 400 eV, and 496 eV was used for interstitial Ag hopping in Al bulk, Ag adatom hopping on Ag island, $As_2$ reaction with Bi-terminated GaAs surface, and $PF_6^-$ diffusing through graphdiyne, respectively. The following projector augmented wave potentials[14] were used: Ag_GW ($5s^1 4d^{10}$) for Ag, Al_GW ($3s^2 3p^1$) for Al, As_GW ($4s^2 4p^3$) for As, Bi ($4s^2 4p^3$) for Bi, Ga_GW ($4s^2 4p^1$) for Ga, H_1.25 ($1s^1$) for passivation H in the GaAs case, C_GW ($2s^2 2p^2$) for C, F ($2s^2 2p^5$) for F, and P_h ($3s^2 3p^3$) for P. The supercells size is about $12 \times 12 \times 12$, $23 \times 23 \times 17$, $16 \times 16 \times 17$, and $19 \times 19 \times 12$ Å$^3$ for interstitial Ag hopping in Al bulk, Ag adatom hopping on Ag island, $As_2$ reaction with Bi-terminated GaAs surface, and $PF_6^-$ diffusing through graphdiyne, respectively. The Brillouin zones are sampled with a spacing less than $2\pi \times 0.04$ Å$^{-1}$ in each periodic direction. All major computational parameters are kept identical while comparing the overall CPU time for different methods. A force-scaling constant (POTIM) of 0.5 is generally used for transition state calculations, but a smaller value of 0.3 and 0.1 has to be used to avoid unphysical distortion for the LI method in the interstitial Ag hopping in Al bulk and $As_2$ reaction with Bi-terminated GaAs, respectively.

## III. RESULTS AND DISCUSSIONS

**A Structural Feature in Real Dynamical Processes**

Out start point is analyzing the major changes in the MEP structures of four distinct dynamical processes predicted by density functional theory (DFT), namely, the hopping of an interstitial Ag in Al bulk (Fig. 2a), hopping of an adatom around the corner of a Ag island (Fig. 2b), reaction of an $As_2$ molecule with a Bi-terminated GaAs surface[15] (Fig. 2c), and diffusion of $PF_6^-$ through 2D material graphdiyne[16] (Fig. 2d). Figure 2e-h show that the chemical bond lengths related to the diffusing or reaction atoms change by less than 0.2 Å in the MEP structures. For instance, the shortest Al-Ag distance in the transition state of the first dynamical process is 2.35 Å, which is only 0.12 Å shorter than that in the initial and final state (Fig. 2e). Given that the MEP structures are the energetically most stable ones along the MEP, such similar chemical bond length with the initial and final structures is understandable.

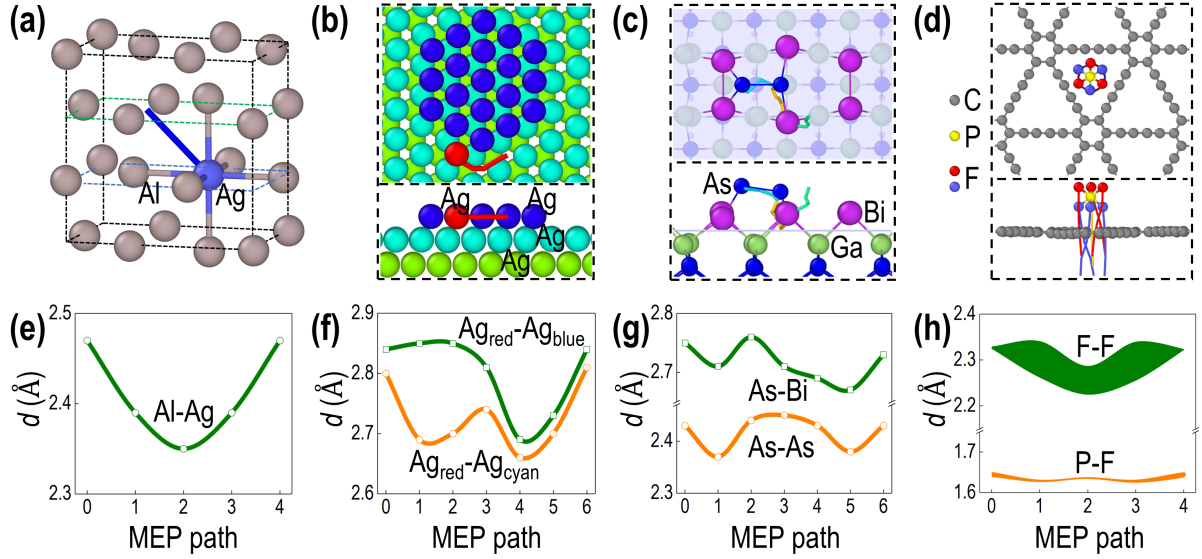

**FIG. 2.** Chemical bond lengths related to the majorly displaced atoms in four DFT-predicted dynamical processes. Initial states of the (a) hopping of an interstitial Ag in Al bulk, (b) hopping of an adatom around the corner of a Ag island, (c) reaction of an $As_2$ molecule with a Bi-terminated GaAs surface, and (d) diffusion of $PF_6^-$ through 2D material graphdiyne; Panels (e)-(h) show the shortest interatomic distances, $d$, related to the majorly displaced atoms, the trajectories of which are indicated by colored tubes in (a)-(d). For easy visualization, a shaded area is plotted in (h) to cover the maximum range of several curves.

**Adaptive Semi-Rigid Body Approximation for Initial Guess of MEP Structures**

Based on the discovery that the chemical bond lengths remain largely unchanged in MEP structures, we propose to approximate the atoms as semi-rigid bodies and assume them in close contact with their nearest atoms in the dynamical process. To obtain the sizes of the semi-rigid bodies or atomic radii for a dynamical process, we utilize the average bond lengths in the initial and final structures and the relationship in Eq. (5)

$$R_i + R_j = \overline{b_{ij}}, \tag{5}$$

where $R_i$ and $R_j$ are the atomic radii of element $i$ and $j$, and $\overline{b_{ij}}$ is their average bond length in the initial and final states. If the number of chemical bond types is more than that of the atom types, one can utilize the singular value decomposition to obtain the optimal values. Relative to the use of empirical atomic radii, the chemical bond lengths in the initial and final states are based on first-principles calculations and thus accurately reflect the chemical and physical differences across materials and dynamical processes. In this sense, the atom radii defined in Eq. (5) are adaptive.

Our next step is selecting atoms that will be adjusted by our proposed method. Because the major structural changes usually occur to a limited number of atoms, such as those around a diffusion path, we select these atoms to optimize their coordinates using the ASRBA method, and the rest can be reliably approximated using the linear interpolation between the initial and final states. Specifically, we compare the initial and final states, and pick out atoms with displacement greater than a threshold, which is set to 0.3 Å in this study, and also their first to third neighboring atoms during a dynamical process (Fig. 3). We label all these atoms as nonlinear atoms, and the rest as linear atoms.

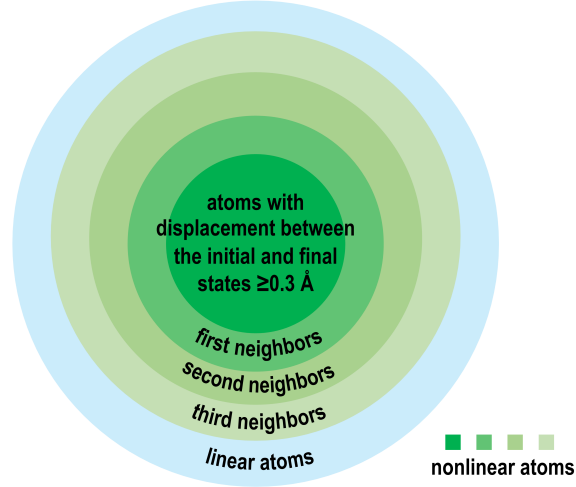

**FIG. 3.** Classification of nonlinear atoms that are optimized using the ASRBA method and the linear atoms that are generated using the LI method.

With the atomic radii and the chosen nonlinear atoms, we propose the following force model to optimize their positions, so a close contact between atoms can be maintained in the guess of the MEP structures. Specifically, if two nonlinear atoms $i$ and $j$ are too close relative to their sum of radii, namely $d_{ij} < (1 - \delta_1)(R_i + R_j)$, they are subjected to a repulsive force as defined by Eqs. (6.1) and (6.2),

$$\boldsymbol{F}_i^{rep} = k^{rep}[d_{ij} - (1 - \delta_1)(R_i + R_j)]^2 / d_{ij}^4 \frac{\boldsymbol{r}_i - \boldsymbol{r}_j}{|\boldsymbol{r}_i - \boldsymbol{r}_j|} \quad (6.1)$$

$$\boldsymbol{F}_j^{rep} = -\boldsymbol{F}_i^{rep} \quad (6.2)$$

, where $k^{rep}$, $d_{ij}$, and $\delta_1$ are a repulsive force constant, the distance between atom $i$ and $j$, and a positive small value, respectively. $\delta_1$ is introduced to allow the atoms to be semi-rigid, as shown by the oscillations in Fig. 2e-h; $d_{ij}^{-4}$ is utilized to reduce the contributions from distant pairs, following the same spirit of Eqs. (3) and (4). By contrast, if two nonlinear atoms are far away but not too far away relative to their sum of radii, namely $(1 + \delta_2)(R_i + R_j) < d_{ij} < (1 + \delta_{cut})(R_i + R_j)$, they are subjected to an attractive force, as defined by Eqs. (7.1) and (7.2),

$$\boldsymbol{F}_i^{att} = k^{att}\big[d_{ij} - (1 + \delta_2)(R_i + R_j)\big]^2 / d_{ij}^{\,4} \frac{r_j - r_i}{|r_i - r_j|} \tag{7.1}$$

$$\boldsymbol{F}_j^{att} = -\boldsymbol{F}_i^{att} \tag{7.2}$$

, where $k^{att}$ is an attractive force constant, and $\delta_2$ and $\delta_{cut}$ are two positive values. For nonlinear atoms with distances too far way, namely $d_{ij} > (1 + \delta_{cut})(R_i + R_j)$, their interactions are ignored. Figure 4 schematically illustrates the forces between nonlinear atoms according to Eqs. (6.1) and (7.1).

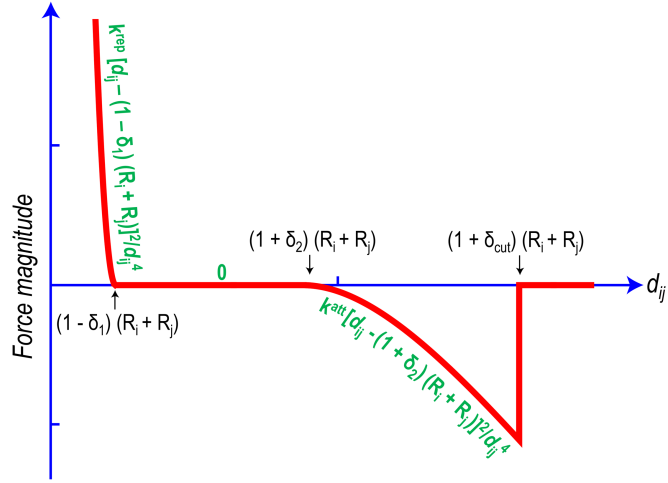

**FIG. 4.** Schematic force between nonlinear atoms $i$ and $j$.

Given that certain systems with π bond, such as graphite and graphdiyne, show anisotropic atomic size, which cannot be included through Eq. (5), we introduce a π bond length $R_\pi$ to Eqs. (6.1) and (7.1) for atomic pairs involving π bond and update them to Eqs. (8.1) and (8.2), respectively.

$$\boldsymbol{F}_i^{rep} = k^{rep}\big[d_{ij} - (1 - \delta_1)(R_i + R_j + R_\pi)\big]^2 / d_{ij}^{\,4} \frac{r_i - r_j}{|r_i - r_j|} \tag{8.1}$$

$$\boldsymbol{F}_i^{att} = k^{att}\big[d_{ij} - (1 + \delta_2)(R_i + R_j + R_\pi)\big]^2 / d_{ij}^{\,4} \frac{r_j - r_i}{|r_i - r_j|} \tag{8.2}$$

Finally, we also allow the linear atoms to impose attractive and repulsive forces to the nonlinear atoms according to Eqs. (6.1), (7.1), (8.1) and (8.2), but these forces are not allowed to impact the linear atoms, whose positions based on the linear interpolation method are supposed to be accurate enough. We further employ the NEB method to maintain an equal distance between the intermediate structures and utilize the above forces as the true forces in the NEB calculation.[4]

Based on the adaptive semi-rigid body approximation of the proposed method, we name it as ASRBA and implement it in the pymatgen code.[8,9] The empirical values of $\delta_1$, $\delta_2$, $\delta_{cut}$, $k^{rep}$, $k^{att}$ and $R_\pi$ are set to be 0.03, 0.10, 0.25, 10 eV·Å, 0.15 eV·Å, and 0.7 Å, respectively.

**Application Example 1: Hopping of Interstitial Ag in Al Bulk**

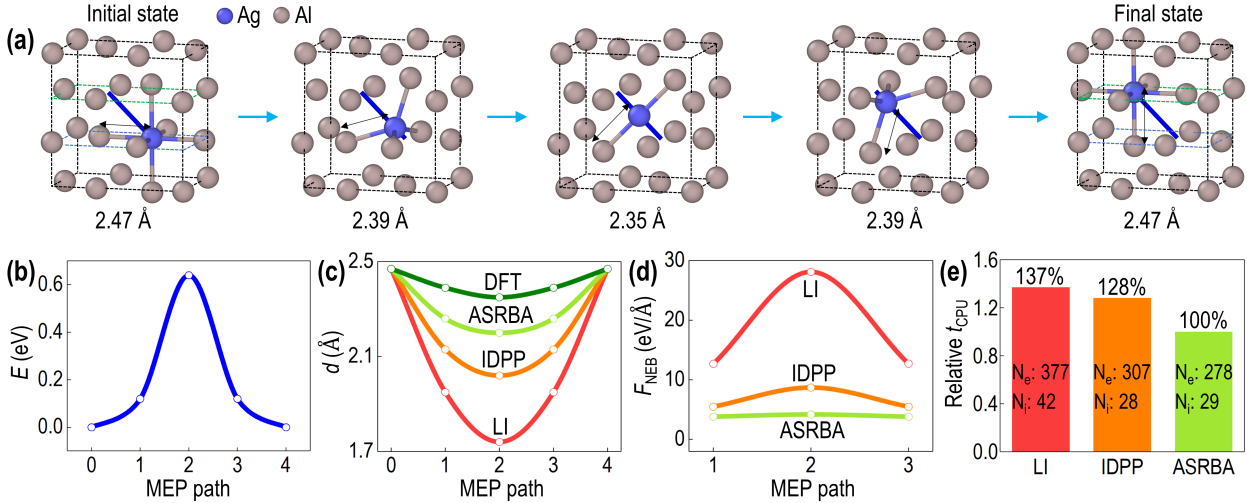

**FIG. 5.** Hopping of interstitial Ag in Al bulk. (a) Optimized trajectory by DFT calculation. The blue tube illustrates the trajectory of Ag atom and the shortest Ag-Al distance is indicated by arrow and labeled below each structure; (b) optimized minimum energy curve versus the MEP path; (c) comparison of the shortest Ag-Al distance among the DFT, LI, IDPP, and ASRBA results; (d) initial NEB forces perpendicular to spring for the LI, IDPP, and ASRBA results; (e) relative CPU time for DFT calculation based on the LI, IDPP, and ASRBA results. $N_e$ and $N_i$ represents the total electronic and ionic steps during the DFT calculations, respectively.

Our first application example is the hopping of an interstitial Ag in Al bulk, which serves as a representative of defect diffusion in bulk. Figure 5a shows the trajectory of the hopping process, where a Ag atom starts at an octahedra center and ends at an equivalent site nearby, and the corresponding energy curve is shown in Fig. 5b. Figure 5c compares the shortest Ag-Al distance in the initial guesses of MEP structures generated by the LI, IDPP, and ASRBA methods relative to the DFT results. It turns out that the largest deviation is 0.61 and 0.33 Å for the LI and IDPP results, respectively, but only 0.15 Å for the ASRBA result. Consequently, the initial forces on the intermediate structures differ significantly among the three methods: the largest initial NEB forces are 28.1, 8.7, and 4.2 eV/Å for the LI, IDPP, and ASRBA results (Fig. 5d), respectively. Note that since the LST method can easily generate discontinuous trajectory, it is not considered in this and following tests. As expected, the subsequent cNEB calculation with the ASRBA result is noticeably faster than the two others, with a relative CPU time of 1.37, 1.28, and 1 for the LI, IDPP, and ASRBA methods, respectively (Fig. 5e). Overall, the ASRBA method outperforms the LI and IDPP method in the initial structural deviation relative to the DFT result, initial NEB forces, and the CPU time of the following DFT calculations.

**Application Example 2: Hopping of Adatom around the Corner of Ag Island**

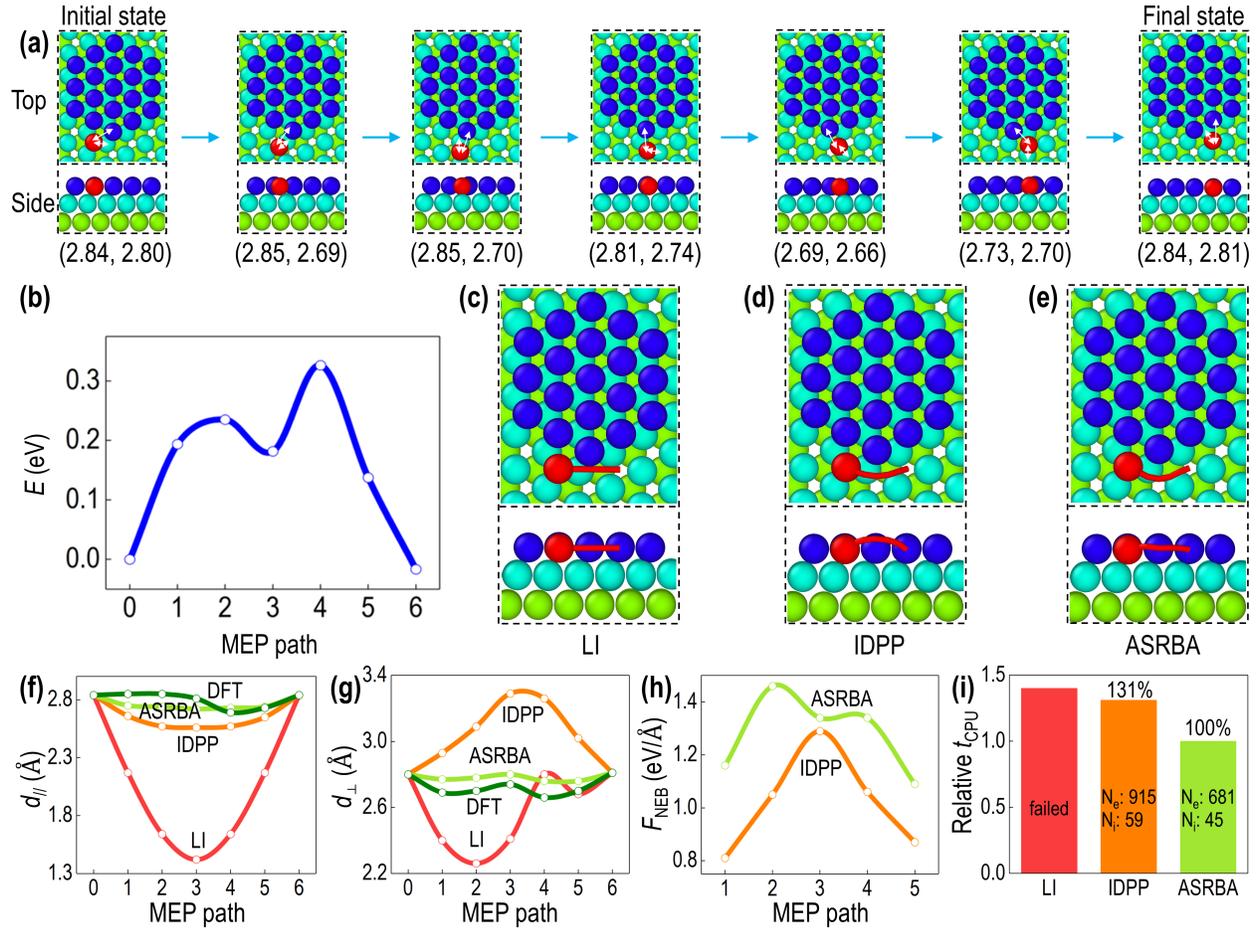

**FIG. 6.** Hopping of a Ag adatom around the corner of a Ag island. (a) Optimized trajectory by DFT calculation, with the hopping Ag adatom in red. Two shortest Ag-Ag distances $d_{//}$ and $d_{\perp}$ related to the hopping atom in the same and different layers, respectively, are indicated by arrows and labeled at the bottom of each structure in the unit of angstrom; (b) optimized minimum energy curve versus MEP path; trajectories of the hopping Ag adatom generated by the (c) LI, (d) IDPP, and (e) ASRBA methods; comparison of (f) $d_{//}$ and (g) $d_{\perp}$ among the DFT, LI, IDPP, and ASRBA results; (h) initial NEB force perpendicular to spring for the IDPP and ASRBA results; (i) relative CPU time for DFT calculation based on the IDPP and ASRBA results. $N_e$ and $N_i$ represents the total electronic and ionic steps, respectively.

The second example is the hopping of a Ag adatom around the corner of a Ag island. The DFT-optimized trajectory is shown in Fig. 6a and the corresponding energy curve is in Fig. 6b. Figure 6c-e compare the trajectories of the hopping atom predicted by the LI, IDPP, and ASRBA methods, respectively. We find that the LI method predicts a trajectory with too close atomic distance around the corner of the island; the IDPP method avoids such issue, but its trajectory shows an unphysical upward curve, which can be explained by Eq. 2 and the nearly-mirror symmetry between the initial and final structures. By contrast, the ASRBA method predicts a trajectory closely resembling the DFT result. Figure 6f, g compare two shortest Ag-Ag distances related to the hopping atom among different methods. The largest deviations of the structures predicted by the LI, IDPP, and ASRBA methods relative to the DFT result are 1.39, 0.28, and 0.11 Å for $d_{//}$ and 0.44, 0.60, and 0.10 Å for $d_{\perp}$, respectively, clearly demonstrating the advantage of the ASRBA method. Interestingly, Fig. 6h shows that the initial NEB forces on the IDPP structures are lower than those on the ASRBA structures, despite that the latter are closer to the DFT result. This phenomenon can be explained by IDPP's unphysical upward pathway, which enters a relatively smooth potential region. Because the LI method predicts structures with too close Ag-Ag distances, its cNEB calculation fails at the beginning. For the ASRBA and IDPP structures, the subsequent cNEB calculation based on the former is 31% faster than that based on the latter (Fig. 6i).

**Application Example 3: Reaction of an $As_2$ Molecule with a Bi-terminated GaAs Surface**

We further examine the ASRBA method with a more complex dynamical process, namely the reaction of an $As_2$ molecule with a Bi-terminated GaAs surface, which is related to the growth of $GaAs_{1-x}Bi_x$.[15] Figure 7a illustrates the dynamical process of an adsorbed $As_2$ molecule switching position of one As atom with a Bi atom in the Bi-terminated surface and the corresponding energy curve is shown in Fig. 7b. As in the two previous examples, the ASRBA method generates intermediate structures closest to the DFT result. The largest deviations of the As-Bi (As-As) distance in the LI, IDPP, and ASRBA results are 1.16 (0.51), 0.27 (0.21), and 0.19 (0.17) Å, respectively, relative to the DFT result, as shown in Fig. 7c (Fig. 7d). Figure 7e indicates that the largest initial NEB force for the LI result is up to 139.6 eV/Å, while that of the IDPP and ASRBA is 4.2 and 3.1 eV/Å, respectively. As expected, the subsequent cNEB calculation with the ASRBA result exhibits the best performance, and the relative CPU time is 2.04, 1.44, and 1 for those based on the LI, IDPP, and ASRBA results, respectively (Fig. 7f).

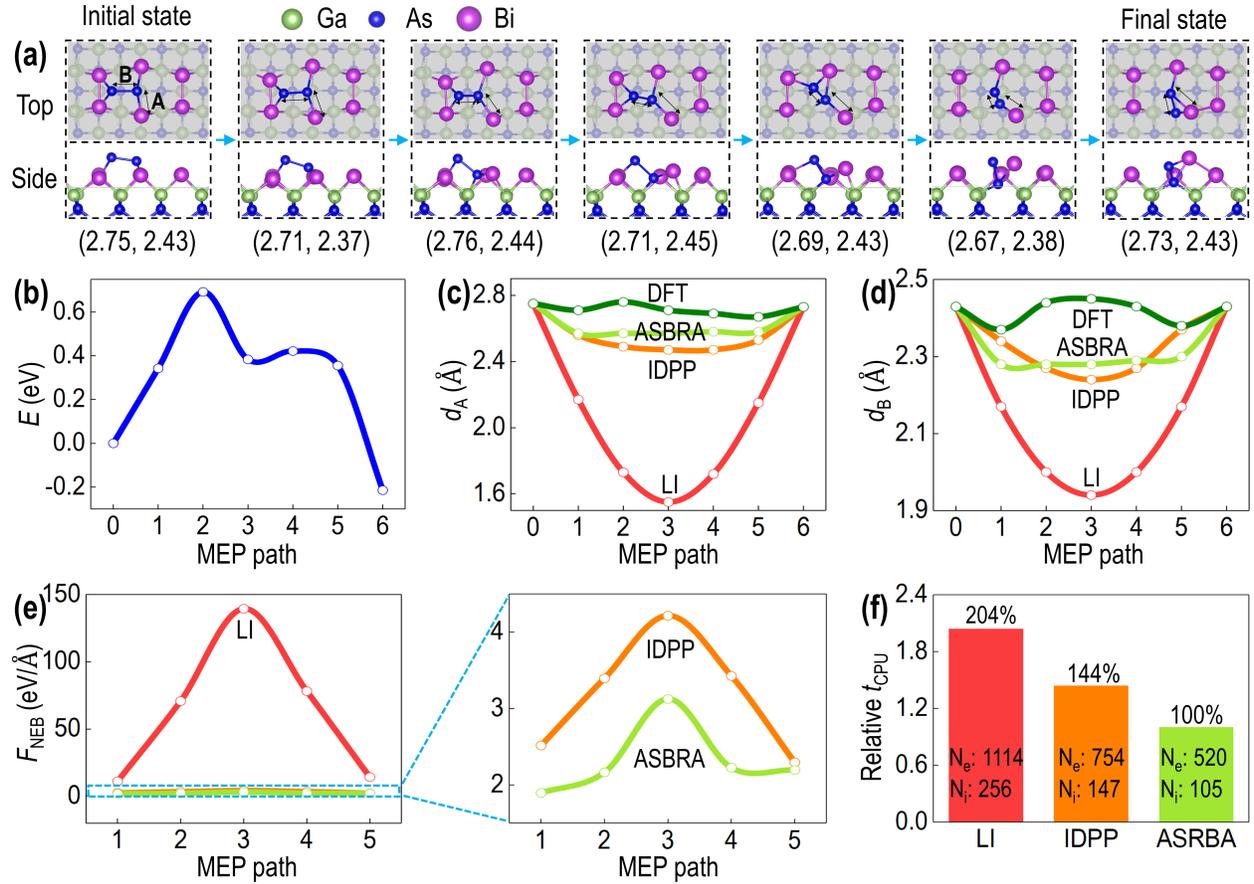

**FIG. 7.** Reaction of an As$_2$ molecule with a Bi-terminated GaAs surface. (a) Optimized trajectory by DFT calculation, with the shortest As-Bi distance $d_A$ and As-As distance $d_B$ indicated by arrows and labeled at the bottom of each structure in the unit of angstrom; (b) optimized minimum energy curve versus the MEP path; comparison of (c) $d_A$ and (d) $d_B$ among the DFT, LI, IDPP, and ASRBA results; (e) initial NEB force perpendicular to spring for the LI, IDPP, and ASRBA results; (f) relative CPU time for DFT calculations based on the LI, IDPP, and ASRBA results. $N_e$ and $N_i$ represents the total electronic and ionic steps, respectively.

**Application Example 4: Diffusion of PF$_6^-$ through 2D Material Graphdiyne**

We finally apply the ASRBA method to the diffusion of PF$_6^-$ through 2D material graphdiyne.[16] Figure 8a shows that PF$_6^-$ passes through a hole in graphdiyne with a ~60° rotation and the corresponding energy curve is shown in Fig. 8b. Unlike the previous three examples, the C atoms in the chains of graphdiyne show anisotropic size, because of the π bond with the acetylenic linkages (-C≡C-).[17] Therefore, we utilize Eqs. (8.1) and (8.2) for the atomic pairs between C and PF$_6^-$. Figure 8c, d compare the lengths of neighboring F-F atoms predicted by the LI, IDPP, and ASRBA methods with the DFT results. The largest deviations of the LI, IDPP, and ASRBA results relative to the DFT result are 0.21-0.24, 0.33-0.36, and 0.07-0.11 Å for the F-F$_{//}$ bonds, and 0.05-0.29, 0.06-0.6, and 0-0.46 Å for the F-F⊥ bonds. Figure 8e, f show similar results for the P-F and C-F bonds, and the largest deviations of the LI, IDPP, and ASRBA results relative to the DFT result are 0.13-0.14, 0.01-0.02, and 0.01-0.06 Å for the P-F bonds, and 0.03-0.60, 0.42-0.62, and 0-0.04 Å for the C-F bonds, respectively. Overall, the ASRBA method predicts a trajectory closest to the DFT result. Figure 8g shows that the largest initial NEB forces in the middle structure are 17.1, 24.2, and 15.0 eV/Å for the LI, IDPP, and ASRBA methods, respectively. Subsequently, the cNEB calculation with the ASRBA result outperforms that of the LI and IDPP results, with relative CPU time of 1.17, 4.17, and 1 for calculations based on the LI, IDPP, and ASRBA results, respectively. Interestingly, we find that the IDPP method is noticeably worse than the LI method, despite its better performance in the previous three examples. This underscores the deficiency of IDPP method for dynamical processes with nearly-mirror symmetry between the initial and final states.

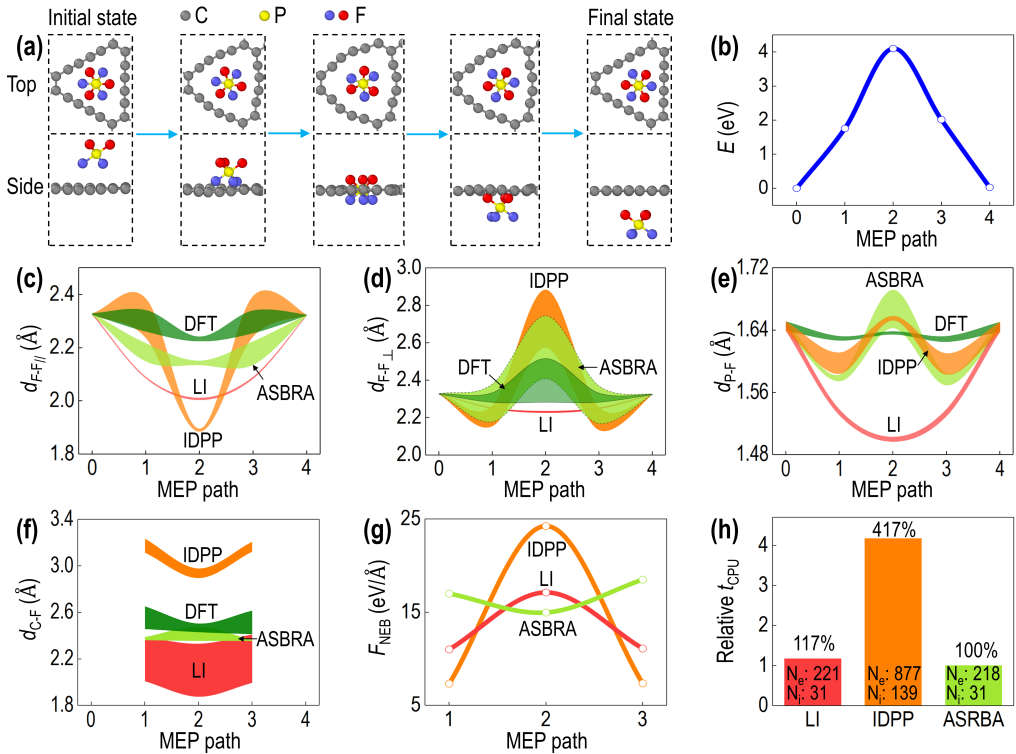

**FIG. 8.** Diffusion of $PF_6^-$ through a 2D material graphdiyne. (a) Optimized trajectory by DFT calculation; (b) optimized minimum energy curve versus the MEP path; comparison of the lengths of (c) F-F bonds parallel to graphdiyne, (d) F-F bonds across graphdiyne, (e) the P-F bonds, and (f) the C-F bonds among the DFT, LI, IDPP, and ASRBA results; (g) initial NEB force perpendicular to spring for the LI, IDPP, and ASRBA results; (h) relative CPU time for DFT calculation based on the LI, IDPP, and ASRBA results. $N_e$ and $N_i$ represents the total electronic and ionic steps, respectively.

## IV. CONCLUSIONS

Through analyses of several distinct dynamical processes predicted by DFT calculations, we find that the chemical bond lengths in the MEP structures largely resemble those in the stable initial and final states. Based on such discovery, we propose the ASRBA method, which extracts the atomic radii from the average bond lengths from the initial and final structures and utilizes a semi-rigid-body force model and NEB method to keep the majorly displaced atoms in close contact within the intermediate structures. Compared to the widely-used LI and IDPP methods, the ASRBA method generates intermediate structures much closer to the DFT results, exhibits noticeably smaller initial NEB forces, and speedups the subsequent cNEB calculations by about 17%–317% for examined dynamical processes in bulk, on crystal surface, and through 2D material. With accelerated transition state calculations, we envision the possibility of creating first-principles dynamics database and further building of machine learning models to predict the MEP structures with unprecedented precisions.

## SUPPLEMENTARY MATERIAL

See the supplementary material for the DFT trajectories of all dynamical processes, and the initial guesses of MEP structures generated by the LI, IDPP, and ASRBA methods.

## ACKNOWLEDGEMENTS

This work was financially supported by the Guangdong Provincial Key Laboratory of Computational Science and Material Design (Grant No. 2019B030301001), the Introduced Innovative R&D Team of Guangdong (Grant No. 2017ZT07C062), and the Shenzhen Science and Technology Innovation Commission (No. JCYJ20200109141412308). The calculations were carried out on the Taiyi cluster supported by the Center for Computational Science and Engineering of Southern University of Science and Technology and also on The Major Science and Technology Infrastructure Project of Material Genome Big-science Facilities Platform supported by Municipal Development and Reform Commission of Shenzhen.

## AUTHOR DECLARATIONS

**Conflict of Interest**

The authors have no conflicts to disclose.

## Author Contributions

**Hongsheng Cai:** Data curation (equal); Formal analysis (equal); Methodology (equal); Software (equal); Writing – original draft (lead). **Guoyuan Liu:** Data curation (equal); Formal analysis (equal); Methodology (equal); Software (equal); Writing – original draft (supporting). **Peiqi Qiu:** Formal analysis (supporting). **Guangfu Luo:** Conceptualization (lead); Formal analysis (equal); Funding acquisition (lead); Methodology (lead); Resources (lead); Supervision (lead); Writing – review & editing (lead).

## DATA AVAILABILITY

The data that support the findings of this study are available from the corresponding author upon reasonable request.